# Major and minor. The formula of musical emotions.
## Vadim R. Madgazin    mailto:vrm@vmgames.com

*The new formulas, which determine sign and amplitude of utilitarian emotions, are proposed on the basis of the information theory of emotions. In area of perception of musical chords the force of emotions depends on the relative pitch of sounds of major and minor chords.*

*Is advanced hypothesis that in the perception of a musical chord in the psyche caused by the subject value of some objective function L. This function is expressed directly through the proportion of the pitch of chord. Major chords are expressed as the straight proportions, which generate idea about an increase in the objective function (L>1) and are caused positive utilitarian emotions.*
*Minor chords are expressed as the inverse proportion, which generate idea about the decrease of objective function (L<1) and are caused negative utilitarian emotions.*

*The formula of musical emotions is advanced: Pwe = log(L) = (1/M)\*log(n1\*n2\*n3\* ... \*nM), where M is a quantity of voices of chord, ni - integer number (or reciprocal fraction) from the pitch proportion, which corresponds to the i-th voice of chord.*

*Confined experimental check is produced. The limits of the applicability of the formula of musical emotions are investigated.*

## Мажор и минор. Формула музыкальных эмоций.
### Вадим Р. Мадгазин

*В соответствии с информационной теорией эмоций в работе предложены модифицированные формулы, выражающие знак и амплитуду утилитарных эмоций через параметры ситуации.*

*Выдвинута гипотеза о том, что при восприятии музыкального аккорда в психике субъекта порождается значение некоторой целевой функции L, которое непосредственно связано с пропорцией высот звуков аккорда. При этом мажорным аккордам соответствуют прямые пропорции, порождающие представление о росте целевой функции (L>1), вызывающее положительные утилитарные эмоции, а минорным аккордам соответствуют обратные пропорции, порождающие представление о падении целевой функции (L<1), вызывающее отрицательные утилитарные эмоции.*

*Выдвинута формула музыкальных эмоций: Pwe = log(L) = (1/M)\*log(n1\*n2\*n3\* ... \*nM), где M - количество голосов аккорда, ni - целое число (или обратная дробь) из общей пропорции высот, соответствующее i-му голосу аккорда.*

*Произведена ограниченная экспериментальная проверка, исследованы границы применимости формулы музыкальных эмоций, в которых она верно передает знак и (на мой взгляд) их амплитуду.*

**Ключевые слова: звук, музыка, аккорд, мажор, минор, эмоции, формула музыкальных эмоций, информационная теория эмоций.**

## ПРЕЛЮДИЯ

Мягко зазвучала негромкая музыка. Ее неспешные минорные аккорды плавно разливались вокруг, унося нас куда-то в глубокую даль. Почему-то повеяло грустью... затем темп стал возрастать, высокие ноты сменялись низкими, напряжение постепенно увеличивалось и вот наконец зазвучала яркая, торжественно-радостная, мажорная развязка. Что с нами было? Загадка природы...

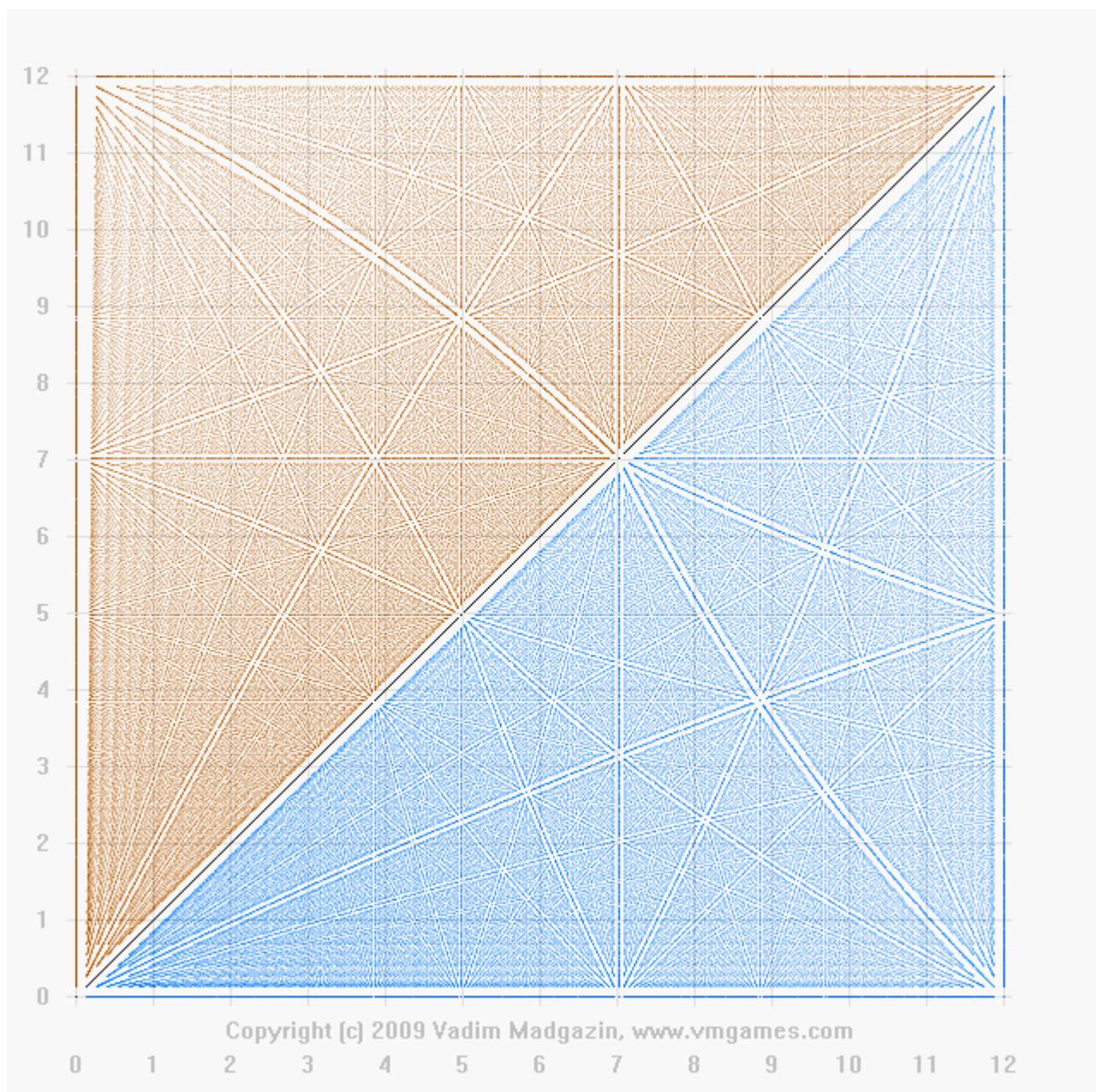

Рисунок 1. Пространство Мажорных и Минорных триад октавы. Цифры на осях - интервалы двух верхних голосов триады относительно нижнего третьего голоса в полутонах РТС12. Показаны только треугольные половинки двух симметричных относительно диагонали пространств, в реальности налагающихся друг на друга.

# ЗВУКИ, АККОРДЫ И АФФЕКТЫ.

Во избежании двусмысленностей приведу несколько вводных фраз, уточняющих терминологию.

Как известно, любой звуковой сигнал ограниченной длительности можно представить в виде эквивалентного ряда (спектра) Фурье как сумму "чистых" тонов (синусоидальных колебаний) с различной амплитудой, частотой и начальной фазой. В данной работе мы будем рассматривать в основном стационарные звуковые сигналы, не меняющиеся во времени.

Согласно [9] основным тоном (первой гармоникой) звука называется наинизшая частота звука. Все остальные частоты выше основного тона называются обертонами. Т.о. первый обертон - это 2-й по высоте тон спектра звука. Обертон с частотой, в N раз большей частоты основного тона (где N - целое число, большее 1) называется N-й гармоникой.

Музыкальным (или гармоническим) называется звук, который состоит только из набора гармоник. На практике это звук, все обертоны которого примерно укладываются на частоты гармоник, а некоторые произвольные гармоники могут отсутствовать, в том числе и первая. В этом случае основной тон называется "виртуальным" и его высота будет определяться психикой субъекта-слушателя из соотношений частот между реальными обертонами [12, 27].

Один музыкальный звук от другого может отличаться по частоте основного тона (высоте), спектру (тембру) и громкости. В данной работы эти отличия не будут использоваться, но все наше внимание будет сконцентрировано на взаимном соотношении высот звуков.

Мы будем рассматривать эффекты от прослушивания одного или нескольких совместных музыкальных звуков, взятых вне любого другого музыкального контекста.

Как известно [4, 9, 16, 27] одновременное звучание двух музыкальных звуков разной высоты (двухголосный аккорд, диада, созвук) способно производить у субъекта впечатление приятного (благозвучного, слитного) или неприятного (раздражающего, шероховатого) сочетания. В музыке это впечатление от созвучий называется соответственно консонансом и диссонансом.

Также известно, что одновременное звучание трех (и более) музыкальных звуков разной высоты (трехголосный аккорд, триада, трезвучие) способно производить у субъекта эмоциональное впечатление различной окраски. Различной - по знаку (положительному или отрицательному) и силе (глубине, яркости, контрасту) соответствующих эмоций.

Эмоции, вызываемые у людей прослушиванием музыки по своему типу в ряду всех известных эмоций относятся к эстетическим (интеллектуальным) и утилитарным эмоциям. О классификации эмоций, в т.ч. музыкальных см. подробнее [21-24].

Например трезвучие из нот "до,ми,соль" (мажорное) и трезвучие из нот "до,ми-бемоль,соль" (минорное) обладают соответственно ярко выраженной "положительной" и "отрицательной" эмоциональной окраской, обычно именуемой как "радость" и "печаль" (или горе, грусть, страдание, сожаление, скорбь, тоска, уныние - согласно [5]).

Эмоциональная окраска аккордов практически не зависит от изменений общей высоты, громкости или тембра составляющих их звуков. В частности, мы услышим практически неизменную эмоциональную окраску у аккордов из довольно тихих чистых тонов.

Забегая вперед отметим, что если какой-то произвольный аккорд может быть определен как минорный или мажорный, то для подавляющего большинства субъектов вызываемые его звучанием эмоции будут утилитарными, т.е. относится к разряду "печаль или радость" (имея отрицательный или положительный знак эмоции). Эмоциональная же сила (яркость эмоции) этого аккорда в общем случае будет зависеть от конкретики ситуации (состояния субъекта-слушателя и структуры аккорда). По существу (в статистическом смысле) можно поставить однозначное соответствие между мажором/минором и эмоциями, ими вызываемыми. И скорее всего именно эмоциональная окраска этих аккордов позволяет "обычным людям" узнавать мажорную или минорную тональность отдельных аккордов.

В [11] такие категории как "высота, тембр, консонанс, диссонанс" отнесены к эстетической системе коммуникаций человека, а "мажор и минор" - к эмоционально-эстетической.

Т.о. резюмируем, что эстетическая компонента звука "приятно-неприятно" (консонанс и диссонанс) возникает у нас при прослушивании двухголосных аккордов, а эмоциональная компонента звука "радость-печаль" (мажор и минор) возникает у нас только при добавлении третьего голоса. Заметим, что аккорды других типов (не мажорных и не минорных) могут не иметь утилитарной компоненты "содержащихся в них" эмоций.

## ПРОПОРЦИИ АККОРДОВ.

Логично сделать предположение, что при восприятии разного числа одновременных музыкальных звуков срабатывает правило перехода количества (1, 2, 3 ...) в качество. Посмотрим, какие новые качества могут при этом появляться.

Еще в глубокой древности было обнаружено, что аккорд из двух (приятных по отдельности) звуков может быть на слух приятным или неприятным (консонантным или диссонантным).
Как было установлено, такой аккорд звучит консонантно, если отношение высот его звуков (с погрешностью скажем в 1% или менее) составляет пропорцию из относительно небольших целых (натуральных) чисел, в частности из чисел от 1 до 6 и 8.
Если же эта пропорция состоит из относительно больших взаимно простых чисел (15/16 и т.п.), то такой аккорд звучит диссонантно.
Отмечу, что точность, с которой следует определять целые пропорции музыкальных звуков как и выбор конкретной пропорции из ряда альтернатив может зависеть от контекста ситуации [2,18]. Краткий исторический экскурс по музыкальным интервалам приведён в [13].

Список отношений высот двух музыкальных звуков (музыкальных интервалов) в порядке уменьшения консонантности согласно [15, 16] выглядит так: 1/1, 2/1, 3/2, 4/3, 5/4, 8/5, 6/5, 5/3, и далее диссонансы 9/5, 9/8, 7/5, 15/8, 16/15.
Этот список возможно не совсем полон (по крайней мере по части диссонансов), т.к. базируется на возможных музыкальных интервалах в рамках равномерно-темперированного строя из 12 нот в октаве (РТС12) .

Известно также, что восприятие консонанса и диссонанса происходит на промежуточном уровне нервной системы человека, на этапе предварительной обработки отдельных сигналов от каждого уха. Если при помощи головных телефонов разделить два звука по разным ушам, то эффекты их "взаимодействия" (пики консонанса, виртуальная высота) исчезают [27,28,29].

Немного отвлекаясь в сторону отмечу, что хотя на сегодняшний день существует более дюжины теорий консонанса и диссонанса [24], дать четкое объяснение почему интервал 7/5 это диссонанс, а 8/5 это консонанс (причем, более совершенный, чем например 5/3) весьма непросто.
Однако нам по большому счёту здесь это и не нужно. Хорошая тема для отдельного исследования?

Итак, отметим следующий новый факт. При переходе от прослушивания одного музыкального звука к двум одновременным звукам у субъекта появляется возможность извлечения информации из соотношения высот этих звуков. Причем, психикой субъекта особо выделяются соотношения высот в виде пропорций из относительно небольших натуральных чисел, которые ранжируются в одной категории - консонанса/диссонанса.

Теперь перейдем к рассмотрению аккордов из трёх звуков. В трезвучиях по сравнению с созвучиями количество (попарных) интервалов увеличивается до трёх, а кроме того появляется новая сущность - само "монолитное" трезвучие (как бы "тройной" интервал) - общее соотношение между высотами всех трёх звуков, рассматриваемых вместе.

Это монолитное соотношение может быть записано как "прямая" пропорция A:B:C или в другой форме как "обратная" пропорция (1/D):(1/E):(1/F) из натуральных взаимно простых троек чисел A,B,C или D,E,F. Чисто математически все такие пропорции можно разделить на три основные группы:
-прямая пропорция "проще" чем обратная, т.е.  $A*B*C < D*E*F$
-обратная пропорция "проще" чем прямая, т.е.  $A*B*C > D*E*F$
-обе пропорции одинаковы ("симметричны"), т.е. $A*B*C = D*E*F$ (и т.о. A=D, B=E, C=F).

Т.о. новое качество трезвучия - информация нового типа - может быть заключена только в этих тройных пропорциях, попадающих в одну из трех вышеописанных категорий.

В зависимости от степени консонантности всех попарных интервалов трезвучия могут быть как консонантными, так и диссонантными. В ряде случаев (при использовании различных целых приближений) выбор конкретного состава обоих пропорций может быть неоднозначным. Однако для консонантных аккордов такая неоднозначность не проявляется.

Согласно [19] в музыкальной практике существуют четыре основных типа трезвучий - мажорное и минорное (консонансы), увеличенное и уменьшенное (диссонансы).
Практически все консонирующие аккорды можно отнести к категории мажора и минора.

Для простоты далее по тексту будем опускать единицу в числителе и круглые скобки для обратных пропорций, т.е. вместо "(1/A)" писать "/A".

Отношения высот звуков вышеупомянутого мажорного трезвучия с большой точностью составляют прямую пропорцию 4:5:6. Отношения высот звуков вышеупомянутого минорного трезвучия с большой точностью составляют обратную пропорцию /6:/5:/4.
Прямые и обратные пропорции увеличенного и уменьшенного трезвучий одинаковы, т.к. они состоят из одинаковых интервалов (4-4 и 3-3 полутона РТС12), и  эти равенства пропорций выглядят как /25:/20:/16  =  16:20:25 и соответственно /36:/30:/25 = 25:30:36.

Соотношение высот звуков мажорных трезвучий всегда более просто (с использованием меньших целых чисел) выражаются в прямых пропорциях, а минорных трезвучий - в обратных пропорциях, и это хорошо известный факт. Уже Джозеффо Царлино (1517-1590 гг.) знал противоположное значение мажорного и минорного аккордов ("Istituzione harmoniche" 1558 г.) [4]. Однако и 450 лет спустя не так легко найти серьезную работу, в которой этот факт широко используется для гармонического анализа или синтеза аккордов. Причиной тому возможно стали упорные, но ошибочные попытки различных авторов объяснить феномен мажора и минора (см. ниже).

На основе простой математики и опытных данных, выдвинем постулат: любой мажорный аккорд (он проще в прямой пропорции) можно превратить в минорный (он проще в обратной пропорции), если вместо прямой пропорции записать обратную из тех же самых чисел. Т.е. если пропорция A:B:C это мажор, то обратная (другая!) пропорция /C:/B:/A это минор. Разумеется любую прямую пропорцию можно (без изменений!) представить в виде обратной, и наоборот. В частности 4: 5: 6 = /15:/12:/10 и /4:/5:/6 = 15: 12: 10.

Суммируя все это можно сделать заключение, что три группы, на которые делятся все пропорции высот трезвучий действительно играют важную роль в музыкальной практике, и соответствуют разделению аккордов на мажорные, минорные и "симметричные" (состоящие из одинаковых интервалов).

Можно задаться вопросом: каково "внутреннее" представление музыкальных трезвучий в психике субъекта? Как им используется информация о вышеупомянутом "новом качестве" трезвучия?

С учетом весьма развитого аппарата слуховой системы человека [9, 10, 12, 27] можно предположить, что хотя представить минорное трезвучие в виде прямой пропорции (15:12:10) высшей нервной системе человека вполне по силам, но также (если не проще) ей по силам представить это же трезвучие в виде обратной пропорции (/4:/5:/6), и "при первом же сравнении" этих пропорций (для определения категории) "отбросить" прямую из-за ее в 15 раз большей сложности (произведение трёх чисел прямой и обратной пропорций равно 1800 против 120).

Будем называть далее главной пропорцией аккорда одну из двух пропорций высот его звуков (прямую или обратную), которая состоит из меньших чисел (в смысле их произведения), другую же пропорцию будем называть побочной. Т.о. главной пропорцией мажорного аккорда всегда будет прямая пропорция, а минорного - обратная пропорция.

И наконец отметим, что хотя вышеупомянутые минорное и мажорное трезвучия состоят попарно из одних и тех же интервалов (4:5, 4:6, 5:6), но имеют противоположную эмоциональную окраску, отсутствующую у любой отдельной пары их звуков.
Единственное же отличие у монолитных трезвучий (минора и мажора) - факт взаимного обращения их главных пропорций.

Логично сделать заключение, что соответствующая новая "эмоциональная" информация аккорда содержится именно в этом последнем свойстве (типе главной пропорции), которое только и может проявиться при сочетании трёх и более звуков, но не может быть обнаружено при сочетании двух (т.к. скажем A:B это абсолютно то же самое, что и /A:/B). Другого источника (эмоциональной) информации, заключенной в трезвучии - просто нет и не может быть (не забываем, что мы рассматриваем стационарные звуки с неизменным спектром). Дополнительным подтверждением этого вывода является то, что у звучания "симметричных" аккордов отсутствует утилитарная компонента эмоций.

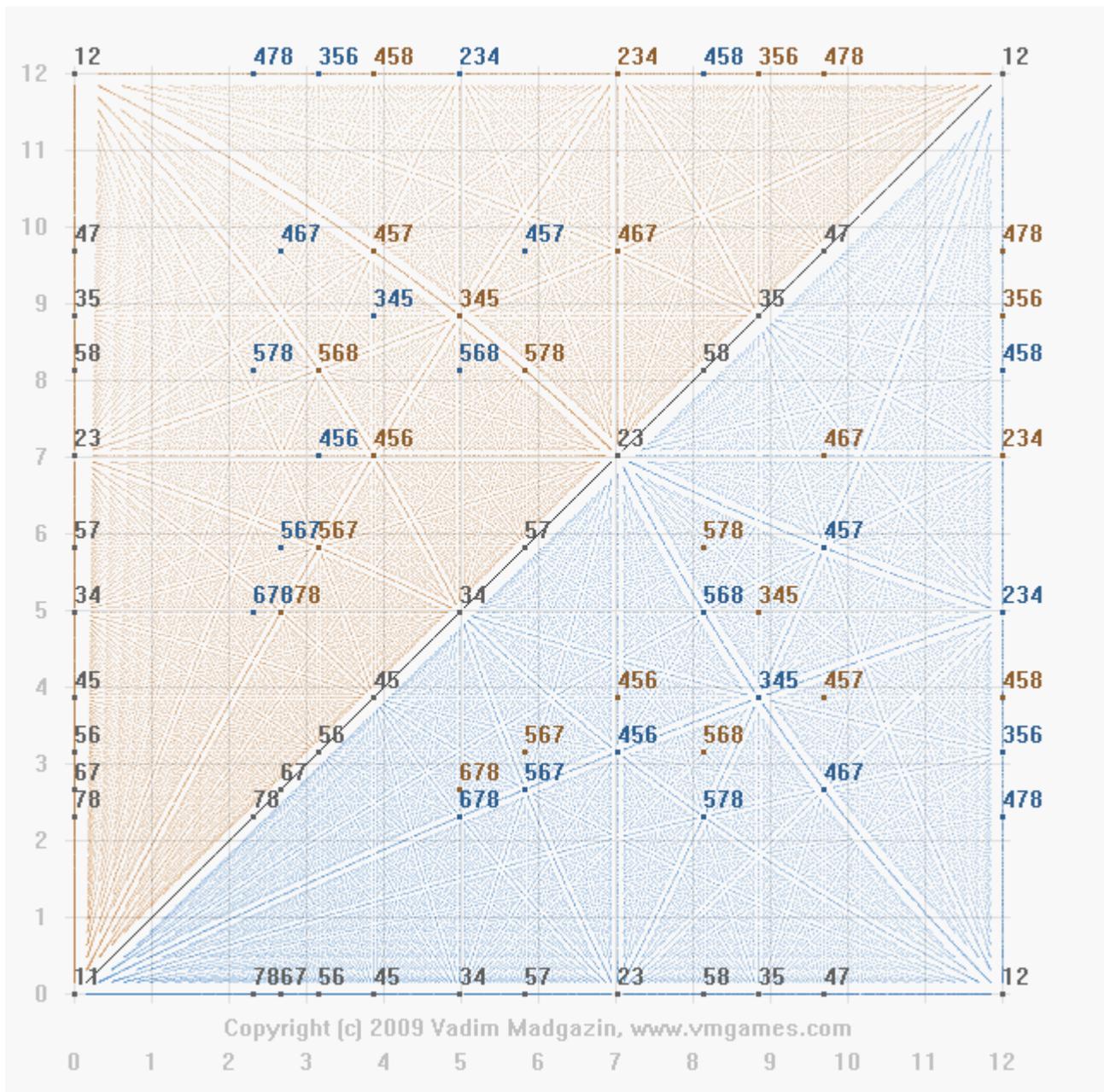

Рисунок 2. Мажорные и Минорные триады и Диады. То же, что и на Рис.1, но с выделением точек простых интервалов и аккордов. Цифры рядом с жирными точками соответствуют числам главных пропорций высот голосов аккордов. Некоторые аккорды диссонантны.

Пример 1. Звучание пропорций.

2:3:4 = /6:/4:/3     даёт мягкий мажор.
2:3:6 = /3:/2:/1     даёт мягкий минор.

3:4:5 = /20:/15:/12  даёт более яркий (контрастный) мажор, а
20:15:12 = /3:/4:/5  даёт более глубокий (контрастный) минор.

4:5:6 = /15:/12:/10  даёт наиболее яркий мажор, а
10:12:15 = /6:/5:/4  даёт наиболее глубокий минор.

Для прослушивания аккордов лучше использовать чистые тона с точными соотношениями частот, используя напр. [1].

## ТЕОРИИ МАЖОРА И МИНОРА.

Аккорды в музыке звучат уже многие сотни лет и почти столько же люди задумываются о причинах их благозвучия.

Для двухголосных аккордов первое объяснение этого свойства было сделано еще очень давно (и подкупающе просто и ясно, если закрыть глаза на некоторые диссонансы - см. выше).
Для трёхголосных аккордов мажора и минора тоже довольно давно были установлены вышеописанные факты о прямых и обратных пропорциях.

Однако найти ответ на вопрос, почему у разных аккордов возникает разная по знаку (и силе) эмоциональная окраска оказалось значительно сложнее. И на второй вопрос - почему минорный аккорд при всей своей сложности (при представлении в прямых пропорциях - так сказать, в "мажорной нотации") звучит благозвучно, а скажем "почти такой же" по сложности числовой пропорции "дисаккорд" (типа 9:11:14) звучит неприятно - ответить было трудно.

Говоря в общем, было не совсем ясно, как "одинаково хорошо" обосновать и мажор, и минор?

Эту загадку природы мажора и минора пытались выяснить многие авторитетные исследователи. И если мажор был-таки объяснён "довольно просто" (как казалось многим авторам, например - "чисто акустически"), то проблема аналогичного по ясности обоснования минора по-видимому до сих пор стоит на повестке дня, хотя и существует великое множество самых разных теоретических и феноменологических построений, пытающихся дать её решение.
Заинтересованный читатель может обратиться за соответствующими подробностями к [2, 3, 4, 13, 16, 24-26].

Исторически теории минора базировались либо на не физических "унтертонах" (призвуках с частотой, в целое число раз меньшей частоты основного тона звука - не существующих в реальности), либо на "метафизических" фактах тройного совпадения обертонов у звуков аккорда, которое хотя и может, но не обязано всегда иметь место - например в случае аккорда из чистых тонов.

Некоторые авторы при "обосновании" аккордов ссылались также на нелинейные свойства слуха, описанного напр. в [9, 10, 12]. Однако и этот бесспорно имеющий место факт весьма редко работает на практике, ведь даже не слишком слабый по громкости аккорд из-за нелинейности не будет порождать различимые комбинационные тоны.

Другие авторы использовали весьма сложные теоретико-музыкальные построения (или чисто математические схемы, замкнутые как "вещи в себе"), разобраться в точном смысле которых часто было невозможно без подробного изучения специфической терминологии самих этих теорий (и порой это объяснение было основано на перефразировке одних абстрактных терминов через другие).

Некоторые авторы до сих пор пытаются подойти к этому вопросу с точки зрения когнитивной психологии, нейродинамики, лингвистики и т.п. И у них почти получается... Почти - потому что цепочка объяснений бывает слишком длинна и далеко не бесспорна, а кроме того отсутствует алгоритмическая формализация теорий и т.о. база для их количественной экспериментальной проверки.

Например в одном из самых интересных, подробных и разносторонних исследований феномена мажора и минора [25] приводится гипотеза о том, что основа эмоционального содержания звуков заложена природой еще в инстинкте высших животных, получившем дальнейшее развитие у человека. Экспериментально установлено, что доминантность конкретной особи стаи в животном мире сопровождается использованием низких или понижающихся звуков "речи", а подчиненность - использованием высоких или повышающихся. Далее принимается, что доминантность равна "радости", а подчиненность - "печали". Затем строится таблица из диссонантных симметричных трезвучных аккордов (с двумя одинаковыми интервалами от 1 до 12 полутонов РТС12) со списком изменений этих аккордов в минорные при увеличении или в мажорные при уменьшении высоты любого звука первоначального аккорда на один полутон.

Даже отвлекаясь от того, что часть измененных аккордов не может быть однозначно отнесена к мажорным или минорным - непонятно, почему слушая аккорд субъект-человек обязательно (и мгновенно) должен "подумать" о том, что один из звуков этого (консонантного) аккорда сдвинут от звука другого (однозначно определенного и к тому же диссонантного) аккорда на некоторый фиксированный интервал - полутон? И как эта довольно абстрактная мысль может превратится во "врожденные" эмоции? И почему разум должен ограничиваться только возможностями РТС12? РТС12 что, тоже придумала Природа и вложила в инстинкт?

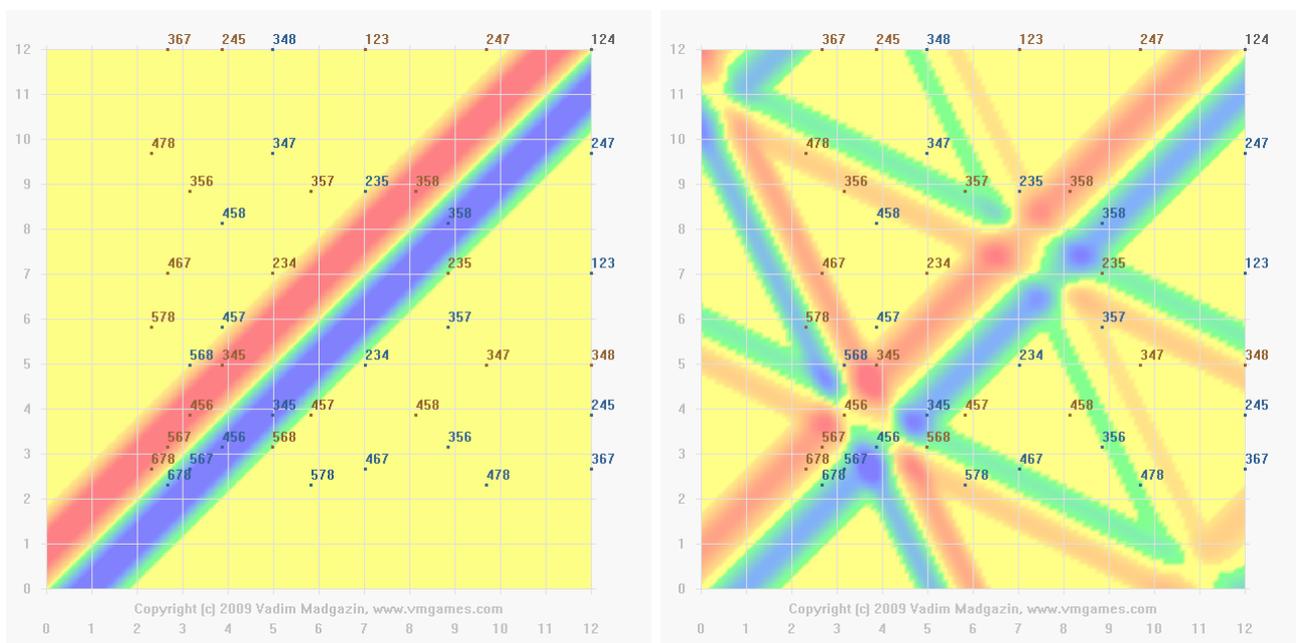

Рисунок 3. По вертикальной оси отложен нижний интервал триад в полутонах РТС12, по горизонтальной оси - верхний интервал. Как и на Рис. 2 показаны точки и пропорции аккордов. Фон заимствован из [30], слева для чистых тонов, справа для тонов из 1-й и 2-й гармоник. Все мажорные аккорды по теории авторов [25,30] должны быть в пределах красного фона, а все минорные - в пределах синего. Желтый фон дает нейтральную область.

На Рис. 3 сопоставлена теория автора [25] с реальностью - набором простых мажорных и минорных триад. Хорошо видно, что множество реальных аккордов выпадают из классификации автора [25], особенно для случая аккордов из чистых тонов (хотя и непонятно, почему тональность аккорда должна так сильно зависеть от его тембра). Многие аккорды "становятся" нейтральными, а некоторые даже "превращаются" в противоположные (например 5:7:8 и 3:5:7).

Тем не менее, я согласен, что эмоциональное содержание мажора и минора основано на эмоциях, доступных многим высшим животным... правда неясно - могут ли они испытывать эти эмоции, прослушивая аккорды? Думаю, вряд ли. Потому что определение взаимных пропорций высот трех и более звуков аккорда - процесс более высокого порядка сложности, чем определение высоты одного звука (или направления изменения этой высоты).

Особое развитие слуховой аппарат человека получил в связи с появлением речевого общения, породившего способность детального и быстрого анализа спектра сложных звуков, побочным продуктом которого скорее всего и является наша возможность наслаждаться музыкой [27].
Утилитарные эмоции у высших животных (как и у человека) однако вполне могут вызываться через восприятие информации от других органов чувств - и прежде всего - через визуальное восприятие событий и их дальнейшую интерпретацию.

Пара слов про эмоциональность речи человека и одноголосной музыки. Да, в них могут "содержаться" утилитарные эмоции. Но причиной тому служит существенная нестационарность спектра - изменения высоты и/или тембра этих звуков.

И еще - про индивидуальные отличия субъектов. Да, при помощи специального воспитания (дрессировки) можно приучить людей (как и некоторых животных) к тому, что даже один какой-то звук (или любой аккорд) будет вызывать у них утилитарные эмоции (горе от рефлекторно ожидаемого кнута или радость - от пряника). Но это не будет соответствовать естественной природе вещей, которую мы стремимся установить.

Вот фраза из докторской диссертации по музыкологии 2008 г., по-видимому ставящая жирную точку в вопросе об известных теориях мажора и минора [26, стр. 91]:
"несмотря на то что многими авторами было описано восприятие мажорных/минорных аккордов и гамм, по-прежнему остается загадкой почему мажорные аккорды ощущаются как радостные, а минорные как грустные".

Я думаю, что разработка верной теории мажора и минора возможна только при удовлетворении двух важных условий:
-привлечении дополнительных областей знаний (кроме музыки и акустики),
-использовании математического аппарата дополнительных областей знаний.

Нам стоит вспомнить историю. Идея о том, что "смысл" аккорда надо искать вне "старого" пространства теории музыки впервые прозвучала как минимум еще сто с лишним лет назад.

Вот пара цитат.
Из [3]:
Гуго Риман (1849-1919) к концу своей деятельности отказался от обоснования мажора и консонанса посредством явления обертонов и встал на психологическую точку зрения Карла Штумпфа, рассматривая обертоны лишь как "пример и подтверждение", но не доказательство.

Из [4, 13]:

Карл Штумпф (1848-1936) перенёс научное обоснование теории музыки из области физиологии в область психологии. Штумпф отказывался объяснять консонанс как акустический феномен, а исходил из психологического факта "слияния тонов" (Stumpf C. Tonpsychologie. 1883-1890).

Итак, завершая раздел отмечу, что скорее всего уже Штумпф и Риман были абсолютно правы в том, что обосновать аккорд невозможно ни акустически, ни метафизически, ни чисто музыкально, а что для этого необходимо - это привлечение психологии.

Теперь подойдём к вопросу "с другого конца" и зададим вопрос: а что же такое есть эмоция?

## ТЕОРИИ ЭМОЦИЙ.

Рассмотрим кратко две теории эмоций, на мой взгляд ближе всего приблизившихся к тому уровню, на котором открывается возможность применения их законов в таком сложном вопросе, как психологическая структура явлений восприятия музыки.

За остальными теориями и подробностями отсылаю читателя к довольно обширному обзору в [6].

Содержание дальнейшего изложения обильно цитирует работы [6, 7, 8] (в моей редакции).

### Фрустрационная теория эмоций.

В 1960-х гг. возникла и была основательно разработана теория когнитивного диссонанса Л.Фестингера.

Согласно этой теории, когда между ожидаемыми и действительными результатами деятельности имеется расхождение (когнитивный диссонанс), возникают отрицательные эмоции, в то время как совпадение ожидания и результата (когнитивный консонанс) приводит к появлению положительных эмоций. Возникающие при диссонансе и консонансе эмоции рассматриваются в этой теории как основные мотивы соответствующего поведения человека.

Несмотря на многие исследования, подтверждающие правоту этой теории, существуют и другие данные, показывающие, что в ряде случаев и когнитивный диссонанс может вызвать положительные эмоции.

По мнению Дж.Ханта, для возникновения положительных эмоций необходима определенная степень расхождения между установками и сигналами, некоторый "оптимум расхождения" (новизны, необычности, несоответствия и т.п.). Если сигнал не отличается от предшествовавших, то он оценивается как неинтересный; если же он отличается слишком сильно, то кажется опасным, неприятным, раздражающим и т.п.

### Информационная теория эмоций.

Несколько позднее оригинальную гипотезу о причинах явления эмоций выдвинул П.В.Симонов.

Согласно ей эмоции появляются вследствие недостатка или избытка сведений, необходимых для удовлетворения потребности субъекта. Степень эмоционального напряжения определяется силой потребности и величиной дефицита прагматической информации, необходимой для достижения цели.

Достоинством своей теории и основанной на ней "формулы эмоций" П.В.Симонов считал то, что она противоречит взгляду на положительные эмоции, как на удовлетворенную потребность. С его точки зрения положительная эмоция возникнет только в том случае, если поступившая информация превысит имевшийся ранее прогноз относительно вероятности удовлетворения потребности.

Многие авторы критиковали формулу эмоций Симонова за ее неполноту, неточность и неопределенность в целом ряде конкретных ситуаций.

Дальнейшее развитие теория Симонова получила в работах О.В.Леонтьева, в частности к 2008 г. в [8] была опубликована весьма интересная статья с рядом обобщенных формул эмоций, одну из которых я ниже подробно и опишу. Далее цитирую [8].

Под эмоциями мы будем подразумевать психический механизм управления поведением субъекта, оценивающий ситуацию по некоторому набору параметров ... и запускающий соответствующую программу его поведения. Кроме того, каждая эмоция имеет специфическую субъективную окраску.

Приведенное определение предполагает, что вид эмоции определяется соответствующим набором параметров. Две различные эмоции должны отличаться различным набором параметров или областью их значений.

Кроме того, в психологии описаны различные характеристики эмоций: знак и сила, время возникновения относительно ситуации - предшествующие (до ситуации) или констатирующие (после ситуации) и т.п. Любая теория эмоций должна давать возможность объективного определения этих характеристик.

Зависимость эмоции от её объективных параметров называют формулой эмоций.

<u>Одно-параметрическая формула эмоций.</u>

Если человек обладает некоторой потребностью величиной П, и если ему удается получить некий ресурс Уд (при Уд > 0), удовлетворяющий потребность, то эмоция Е будет положительной (а в случае потери Уд < 0 и эмоция будет отрицательной):

Е = F(П, Уд)   (1)

Ресурс Уд определяется в работе [8] как "Уровень достижений", а эмоция Е - как констатирующая.
Для конкретности можно представить человека, играющего в новую для себя игру и не представляющего, что можно от нее ожидать.

Радость.
Если игрок выиграл некую сумму Уд > 0, то возникает положительная эмоция радости силой Е = F(П, Уд).

Горе.
Если игрок "выиграл" сумму Уд < 0 (т.е. проиграл), то возникает отрицательная эмоция горя силой E = F(П, Уд).

Еще один метод формализации эмоций предложен в работе [17].

Согласно ему эмоции рассматриваются как средство оптимального управления поведением, направляющее субъекта к достижению максимума его "целевой функции" L.

Увеличение целевой функции L сопровождается положительными эмоциями, уменьшение - отрицательными эмоциями.

Поскольку L зависит в простейшем случае от некоторой переменной x, то эмоции E вызываются изменением этой переменной от времени:

$E = dL/dt = (dL/dx)*(dx/dt)$   (2)

В [6, 20] также отмечается, что наряду с вышеописанными (утилитарными) эмоциями существуют еще т.н. "интеллектуальные" эмоции (удивление, догадка, сомнение, уверенность и т.п.), которые возникают не в связи с потребностью или целью, а в связи с самим интеллектуальным процессом обработки информации. Например, они могут сопровождать процесс наблюдения абстрактных математических объектов [5]. Особенностью интеллектуальных эмоций является отсутствие у них специфического знака.

На этом этапе остановимся в цитировании и перейдем в основном к изложению оригинальных идей автора.

## МОДИФИКАЦИЯ ФОРМУЛ ЭМОЦИЙ.

Прежде всего отметим, что формулы (1, 2) весьма похожи, если учесть что ресурсный параметр Уд на самом деле является разностью между текущим и предыдущим значением некого интегрального ресурса R. Например, в случае нашего азартного игрока в качестве R логично выбрать его суммарный капитал, тогда:

$Уд = R1 - R0 = dR = dL$

Однако обе формулы (1, 2) "не совсем" физичны - в них приравниваются величины, имеющие разные размерности. Нельзя же измерять, скажем, время в километрах или радость в литрах.

Поэтому во-первых формулы эмоций следует модифицировать, записав их в относительных величинах.
Также желательно уточнить зависимость силы эмоций от своих параметров т.о. чтобы увеличить правдоподобность результатов для большого диапазона изменений этих параметров.

Для этого воспользуемся аналогией с общеизвестным законом Вебера-Фехнера [14], говорящим о том, что дифференциальный порог восприятия для самых разных сенсорных систем человека пропорционален интенсивности соответствующего стимула, а величина

**13**

ощущения пропорциональна его логарифму.

В самом деле, радость того самого игрока должна быть пропорциональна относительному размеру выигрыша, а не абсолютному. Ведь миллиардер, проигравший один миллион будет горевать совсем не так же сильно, как обладатель миллиона с небольшим хвостиком.
И высоты "самых похожих" музыкальных звуков связаны октавным соотношением, т.е. тоже логарифмическим (увеличением частоты основного тона звука в 2 раза).

Я предлагаю записать модифицированную формулу эмоций (1) следующим образом:

$E = F(П) * k * \log(R1/R0),$ (3)

где $F(П)$ - вынесенная отдельно зависимость эмоций от параметра потребности П;
k - некоторая постоянная (или почти постоянная) положительная величина, зависящая от предметной области ресурса R, от основания логарифма, от интервала времени между измерениями R1 и R0, а также возможно от деталей характера конкретного субъекта;
R1 - значение целевой функции (суммарного полезного ресурса) в текущий момент времени,
R0 - значение целевой функции в предыдущий момент времени.

Можно также выразить новую формулу эмоций (3) через безразмерную величину $L = R1/R0$, которую логично назвать относительной дифференциальной целевой функцией (текущим значением интегральной целевой функции относительно некоторого предыдущего момента времени, всегда находящегося на фиксированном расстоянии от момента текущего).

Тогда:

$E = F(П) * Pwe,$ где $Pwe = k * \log(L),$ (4)

где в свою очередь $L = R1/R0$, а параметры k, R0 и R1 описаны в формуле (3).

Здесь введена величина мощности эмоций Pwe, пропорциональная "потоку эмоциональной энергии" в единицу времени (т.е. обиходному смыслу выражения "интенсивность эмоций", "сила эмоций"). Выражение силы эмоций в единицах мощности, выделяемой организмом субъекта на эмоциональное поведение известно из работ других авторов [5], поэтому нам не стоит удивляться появлению такого (несколько необычного) термина как "мощность эмоции".

Как легко видеть, формулы (3 и 4) автоматически дают верный знак эмоций, положительный при росте R (когда $R1 > R0$ и т.о. $L > 1$) и отрицательный при падении R (когда $R1 < R0$ и т.о. $L < 1$).

Теперь попробуем применить новые формулы эмоций к восприятию музыкальных аккордов.

## ИНФОРМАЦИОННАЯ ТЕОРИЯ АККОРДОВ.

Но сначала немного "лирики". Как можно изложить вышеописанную информационную теорию эмоций на простом человеческом языке? Попробую привести несколько довольно незамысловатых примеров, проясняющих ситуацию.

Допустим, сегодня нам жизнь отвалила "двойную порцию" неких "жизненных благ" (против среднего ежедневного объема "счастья"). Например - в два раза лучший обед. Или нам выпало вечером два часа свободного времени против одного. Или мы в горном походе прошли вдвое больше чем обычно. Или нам сказали в два раза больше комплиментов чем вчера. Или мы получили двойные премиальные. И мы радуемся, потому что функция L сегодня стала равной 2 (L=2/1, E>0). А завтра нам все это выпало в пятикратном размере. И мы радуемся еще больше (испытываем более мощные положительные эмоции, потому что L=5/1, E>>0). А потом все это пошло как обычно (L=1/1, E=0), и мы больше не испытываем никаких утилитарных эмоций - нам нечему радоваться, и нечему печалиться (если мы еще не успели привыкнуть к счастливым денькам). А потом вдруг разразился кризис и наши блага урезали наполовину (L = 1/2, E<0) - и нам стало грустно.

И хотя у каждого субъекта целевая функция L зависит от большого набора индивидуальных под-целей (порой диаметрально противоположных - для спортивных противников или болельщиков, например), общим для всех является личное мнение каждого - приближает ли данное событие к каким-то из его целей, или отдаляет от них.

А теперь вернемся к нашей музыке.

Исходя из проверенных фактов науки [14] логично предположить, что при одновременном прослушивании нескольких звуков психика субъекта пытается извлечь всевозможную информацию, которую эти звуки могут заключать в себе, в том числе находящуюся на самом высоком уровне иерархии, т.е. из соотношений высот всех звуков.

На этапе анализа параметров трезвучий (в отличии от созвучий, см. выше) индивидуальные потоки информации от разных ушей используются уже совместно (что легко проверить, подавая любые два звука в одно ухо, а третий - в другое - эмоции те же).

В процессе интерпретации этой объединенной информации психика субъекта пытается использовать в том числе и свою "утилитарную" эмоциональную подсистему.

И в ряде случаев ей это с успехом удается - например, при прослушивании изолированных минорных и мажорных аккордов (но аккорды других типов по-видимому могут порождать другие виды эмоций -  эстетические/интеллектуальные).

Возможно, некие довольно простые аналогии (на уровне больше/меньше) со смыслом "похожей" информации из других сенсорных каналов восприятия (визуальных и др.) позволяют психике субъекта классифицировать мажорные аккорды как несущие информацию "о выгоде", сопровождаемую положительными эмоциями, а минорные - "об убытке", сопровождаемую отрицательными.

Т.е. на языке формулы эмоций (4) в мажорном аккорде должна быть заключена информация о значении целевой функции $L > 1$, а в минорном - о значении $L < 1$.

<u>Моя основная гипотеза состоит в следующем. При восприятии отдельного музыкального аккорда в психике субъекта порождается значение целевой функции L, которое непосредственно связано с главной пропорцией высот его звуков. При этом мажорным аккордам соответствует представление о росте целевой функции (L>1), сопровождаемое положительными утилитарными эмоциями, а минорным аккордам соответствует представление о падении целевой функции (L<1), сопровождаемое отрицательными утилитарными эмоциями.</u>

А первом приближении можно предположить, что значение L равно некоторой простой функции от чисел, входящих в главную пропорцию аккорда. В простейшем случае эта функция может быть каким-то "средним" от всех чисел главной пропорции аккорда, например средним геометрическим.

Для любых мажорных аккордов все эти числа будут больше 1, а для любых минорных они будут меньше 1.

Например:
L =   N = "среднее" от чисел (4, 5, 6)      из мажорной пропорции 4: 5: 6,
L = 1/N = "среднее" от чисел (1/4, 1/5, 1/6) из минорной пропорции /4:/5:/6.

При таком представлении L амплитуда силы эмоций (т.е. абсолютное значение Pwe), порождаемых мажорным и (обратным ему) минорным трезвучием будет совершенно одинакова, и эти эмоции будут иметь противоположный знак (мажор - положительный, минор - отрицательный). Весьма обнадеживающее следствие!

Попробуем теперь уточнить и обобщить формулу (4) для произвольного количества голосов аккорда M. Для этого определим L как среднее геометрическое от чисел из главной пропорции аккорда, получив в итоге окончательный вид "формулы музыкальных эмоций":

$Pwe = k * \log(L) = k * (1/M) * \log(n_1 * n_2 * n_3 * ... * n_M)$,   (5)

где k - по прежнему некоторая положительная константа - см. (3),
а $n_i$ - это i-й голос аккорда в виде соответствующего целого числа или обратной дроби из главной пропорции высот голосов аккорда.

Назовем величину Pwe (из формулы 5) "эмоциональной мощностью" аккорда (или просто мощностью), положительной для мажора и отрицательной для минора (аналогия: поток жизненных сил, у мажора - приток, у минора - отток).

Для единообразия с логарифмической шкалой частот (вспомним об октаве), будем использовать в формуле (5) логарифм по основанию 2. В таком случае можно положить k = 1, т.к. при этом численное значение Pwe будет находится во вполне приемлемом диапазоне вблизи области "единичной" амплитуды эмоций.

Для дальнейшего анализа нам наряду с "главной" может понадобиться также "побочная" мощность аккорда, соответствующая подстановке в формулу (5) его побочной пропорции (см. выше). Если не указано - ниже везде используется "главная" Pwe.

В приложении к статье приведены значения главных и побочных мощностей некоторых аккордов.

## ОБСУЖДЕНИЕ РЕЗУЛЬТАТОВ.

Итак, выдвинув ряд довольно простых и логичных предположений, мы получили новые формулы (3, 4, 5), которые связывают обобщенные параметры ситуации (или конкретные параметры аккордов для формулы 5) со знаком и силой вызываемых ими (в контексте ситуации) утилитарных эмоций.

Как можно оценить этот результат?

Цитирую работу [5]:
"Попыток объективного определения силы эмоции, вероятно, не было. Однако можно предположить, что такое определение должно быть основано на энергетических представлениях. Если эмоция вызывает некоторое поведение, то это поведение требует определенного расхода энергии. Чем сильнее эмоция, тем интенсивнее поведение, тем больше требуется энергии в единицу времени.
Т.е. силу эмоции можно попытаться отождествить с величиной мощности, которую организм выделяет на соответствующее поведение."

Попробуем максимально критически подойти к новому результату, раз его пока не с чем сравнивать.

Во первых, мощность эмоций Pwe из формул (4, 5) хотя и пропорциональна "субъективной силе" эмоций, но их связь может быть не линейной. И эта связь - лишь некая средняя зависимость по всему континууму субъектов, т.е. может быть подвержена значительным (?) индивидуальным отклонениям. Например, "константа" k всё же может изменяться, хотя и не слишком сильно. Возможно также, что вместо среднего геометрического в формуле (5) следует применить какую-то другую функцию.

Во-вторых, если иметь ввиду конкретный вид формулы музыкальных эмоций (5), то следует заметить, что хотя формально в ней M и может быть равно 1 или 2, мы можем говорить о возникновении утилитарных эмоций только при M >= 3. Однако уже при M = 2 возможно наличие эстетических/интеллектуальных эмоций, а при M > 3, есть вероятность появления дополнительных факторов (?), как-то влияющих на результат.

В третьих, по-видимому область валидных значений амплитуды Pwe для категории мажора и минора имеет верхнюю границу 2.7 ... 3.0, но где-то уже со значения 2.4 начинается область насыщения утилитарно-эмоционального восприятие аккордов, и примерно там же проходит нижняя граница диапазона возможного "вторжения" диссонансов.
Но это последнее - скорее общая проблема "не монотонности" ряда диссонантных интервалов, не связанная напрямую с эмоциональным восприятием аккордов. А ограниченность динамического диапазона мощности эмоций - общее свойство любой сенсорной системы человека, легко объясняемое отсутствием аналогий с событиями в "реальной жизни", которые соответствуют слишком быстрым изменениям целевой функции (в 7-8 раз и более).

В четвёртых, "симметричные" (или почти симметричные) аккорды, у которых прямые и обратные пропорции состоят из одинаковых чисел (даже при отсутствии в них явных диссонансов) по-видимому выпадают из нашей классификации - их утилитарно-эмоциональная окраска практически отсутствует, соответствуя случаю Pwe = 0.

Однако можно дополнить формальный результат применения формулы (5) простым полу-эмпирическим правилом: если главная и побочная мощности какого-то аккорда (почти) совпадают по амплитуде, то результатом формулы (5) будет не главная мощность, а полусумма мощностей, т.е. (примерно) 0.
И это правило начинает работать уже при отличии амплитуд главной и побочной Pwe, меньшем чем 0.50.

Скорее всего тут имеет место весьма простое явление: раз невозможно отличить по сложности прямую и обратную пропорцию аккорда, то классификация этого аккорда в категориях утилитарных эмоций ("печали и радости") просто не производится. Однако эти аккорды (как и интервалы) могут порождать эстетические/интеллектуальные эмоции, напр. "удивление", "вопрос", "раздражение" (при наличии диссонансов) и др.

При всех своих мнимых или действительных недостатках формула (5) (и по-видимому, формулы 3 и 4) все же дает нам весьма неплохой теоретический материал для численных оценок силы эмоций.

По крайней мере в одной конкретной области - области эмоционального восприятия мажорных и минорных аккордов.

Давайте попробуем проверить эту формулу (5) на практике, при помощи сравнения пары разных мажорных и минорных аккордов. Весьма неплохой пример - аккорды 3:4:5 и 4:5:6 и их минорные варианты.

Для чистоты эксперимента следует сравнивать пары аккордов, составленных из чистых тонов при их примерно одинаковой средней по уровню громкости, причем у обоих аккордов лучше использовать такие высоты тонов, чтобы "средневзвешенная" частота этих аккордов (в Герцах) была одинакова.

Пара мажорных трезвучий может состоять из тонов частотой напр. 300, 400, 500 Гц и 320, 400, 480 Гц.

На слух мне кажется вполне заметным, что эмоциональная "яркость" мажора 3:4:5 (с Pwe = 1.97) действительно несколько меньше чем у мажора 4:5:6 (с Pwe = 2.30).

Примерно то же самое по моему мнению происходит и с минором /3:/4:/5 и /4:/5:/6.

Это впечатление верной передачи мощности эмоций формулой (5) сохраняется и при прослушивании тех же самых аккордов, составленных из звуков с богатым гармоническим спектром.

## ВЫВОДЫ.

Целью моей работы являлось установление возможной природы эмоциональной окраски музыкальных аккордов, в особенности консонантных мажорных и минорных трезвучий.

В связи с этим были проанализированы некоторые существующие теории эмоций. Для дальнейшего использования я выбрал информационную теорию эмоций, базирующуюся на предположении о том, что положительные эмоции связаны с ростом некоторой целевой функции субъекта, а отрицательные - с её падением.

Соответствующие формулы эмоций мной были модифицированы введением логарифма относительной целевой функции вместо её производной, что по моему мнению сделало их более соответствующими реальности в качественном и количественном отношении.

На основе модифицированных формул эмоций я предложил информационную теорию музыкальных аккордов и конкретную формулу музыкальных эмоций (5), определяющую знак и амплитуду эмоциональной окраски аккордов через параметры взаимной относительной высоты составляющих их звуков.

В работе определены пределы применимости формулы музыкальных эмоций, в которых она верно передает знак и (на мой взгляд) их амплитуду.

Однако для практической проверки выдвинутых в работе идей необходимо выполнить статистическое исследование соответствия реальности полученных на их основе результатов.

В частности для проверки формулы музыкальных эмоций требуется статистически достоверное определение знаков и амплитуд эмоций для максимально большого множества консонантных мажорных и минорных аккордов.

## КОДА

Радостно звучат фанфары!
Затем все встают - и взявшись за руки - a cappella поют Гимн Разуму!
Многовековая тайна мажора и минора наконец разгадана! Мы победили...

## ЛИТЕРАТУРА И ССЫЛКИ.

1. Звуковая система Audiere, http://audiere.sourceforge.net/
   Скачать архив http://prdownloads.sourceforge.net/audiere/audiere-1.9.4-win32.zip?download
   Использовать wxPlayer.exe из папки bin.

2. Трусов В.Н. 2004-2009 Материалы сайта http://www.muzhar.ru/

3. Мазель Л. Функциональная школа. 1934
   (Рыжкин И., Мазель Л., Очерки по истории теоретического музыкознания)

4. Риман Г. Музыкальный словарь (компьютерный вариант). 2004

5. Леонтьев В.О. Десять нерешенных проблем теории сознания и эмоций. 2008

6. Ильин Е.П. Эмоции и чувства. 2001

7. Симонов П.В. Эмоциональный мозг. 1981

8. Леонтьев В.О. Формулы эмоций. 2008

9. Алдошина И., Приттс Р. Музыкальная акустика. 2006

10. Алдошина И. Основы психоакустики. Подборка статей с сайта http://www.625-net.ru

11. Морозов В.П. Искусство и наука общения. 1998

12. Альтман Я.А. (ред.) Слуховая система. 1990

13. Лефевр В.А. Формула человека. 1991

14. Шиффман Х.Р. Ощущение и восприятие. 2003

15. Теплов Б.М. Психология музыкальных способностей. 2003

16. Холопов Ю.Н. Гармония. Теоретический курс. 2003

17. Голицын Г.А., Петров В.М. Информация - поведение - творчество. 1991

18. Гарбузов Н.А. (ред.) Музыкальная акустика. 1954

19. Римский-Корсаков Н. Практический учебник гармонии. 1937

20. Леонтьев В.О. Что такое эмоция. 2004

21. http://ru.wikipedia.org/wiki/Эмоции

22. Klaus R. Scherer, 2005. What are emotions? And how can they be measured?
    Social Science Information, Vol 44, no 4, pp. 695-729

23. BEHAVIORAL AND BRAIN SCIENCES (2008) 31, 559-621
    Emotional responses to music: The need to consider underlying mechanisms

24. Music Cognition at the Ohio State University    http://csml.som.ohio-state.edu/home.html
    Music and Emotion    http://dactyl.som.ohio-state.edu/Music839E/index.html

25. Norman D. Cook, Kansai University, 2002. Tone of Voice and Mind:
    The connections between intonation, emotion, cognition and consciousness.

26. Bjorn Vickhoff. A Perspective Theory of Music Perception and Emotion.
    Doctoral dissertation in musicology at the Department of Culture, Aesthetics and Media,
    University of Gothenburg, Sweden, 2008

27. Terhardt E. Pitch, consonance, and harmony. Journal of the Acoustical Society of America,
    1974, Vol. 55, pp. 1061-1069.

28. ВОЛОДИН А.А. Автореферат докторской диссертации.
    ПСИХОЛОГИЧЕСКИЕ АСПЕКТЫ ВОСПРИЯТИЯ МУЗЫКАЛЬНЫХ ЗВУКОВ

29. Levelt W., Plomp R. The appreciation of musical intervals. 1964

30. Norman D. Cook at al. Seeing Harmony. http://www.res.kutc.kansai-u.ac.jp/~cook

## БЛАГОДАРНОСТИ.

## АВТОРСКАЯ ИНФОРМАЦИЯ.

Обратная связь.

Любая конструктивная критика и замечания к работе будут приняты с благодарностью по email адресу: vrm@vmgames.com

Последняя версия этой и другие работы автора: http://www.vmgames.com/ru/texts/

Версия.

Версия текста от 01 сентября 2009 г.

## ПРИЛОЖЕНИЕ.

Эмоциональная мощность Pwe главных пропорций некоторых аккордов, рассчитанная по формуле (5).

Основная часть пропорций - прямые пропорции, соответствующие мажорным аккордам.
Минорные аккорды могут быть получены из пропорций, зеркально обратных мажорным путем простого изменения знака Pwe главной пропорции (как в паре примеров).
В скобках дана побочная мощность некоторых аккордов, если она по амплитуде приближается к главной.
Для симметричных аккордов обе эти мощности отличаются только знаком.

| Главная пропорция | Побочная пропорция | Pwe главной (побочной) пропорции | Примечание |
|---|---|---|---|
| Некоторые симметричные [псевдо]аккорды | | | |
| 1:1:1 | 1:1:1 | 0  (0) | |
| 1:2:4 | /4:/2:1 | 1  (-1) | |
| 4:6:9 | /9:/6:/4 | 2.58  (-2.58) | "квинтовое" трезвучие |
| 16:20:25 | /25:/20:/16 | 4.32  (-4.32) | увеличенное трезвучие |
| Некоторые консонантные 3-х голосные [псевдо]аккорды | | | |
| 1:2:3 | /6:/3:/2 | 0.86  (-1.72) | |
| 2:3:4 | /6:/4:/3 | 1.53  (-2.06) | |
| 2:3:5 | /15:/10:/6 | 1.64 | |
| 2:3:8 | /12:/8:/3 | 1.86 | |
| 2:4:5 | /10:/5:/4 | 1.77 | |
| 2:5:6 | /15:/6:/5 | 1.97 | |
| 2:5:8 | /20:/8:/5 | 2.11 | |
| 3:4:5 | /20:/15:/12 | 1.97 | |
| /3:/4:/5 | 20:15:12 | -1.97 | |
| 3:4:6 | /4:/3:/2 | -1.53 (2.06) | |
| 3:4:8 | /8:/6:/3 | 2.19 (-2.39) | почти симметрично |
| 3:5:6 | /10:/6:/5 | 2.16 (-2.74) | |
| 3:5:8 | /40:/24:/15 | 2.30 | |
| 3:6:8 | /8:/4:/3 | 2.39 (-2.19) | почти симметрично |
| 4:5:6 | /15:/12:/10 | 2.30 | мажорное трезвучие |
| /4:/5:/6 | 15:12:10 | -2.30 | минорное трезвучие |
| 4:5:8 | /10:/8:/5 | 2.44 (-2.88) | |
| 5:6:8 | /24:/20:/15 | 2.64 | |

Некоторые диссонантные трезвучия

| | | | |
|---|---|---|---|
| 4:5:7 | /35:/28:/20 | 2.38 | |
| 5:6:7 | /42:/35:/30 | 2.57 | |

Некоторые консонантные 4-х голосные [псевдо]аккорды

| | | |
|---|---|---|
| 1:2:3:4 | /12:/6:/4:/3 | 1.15 |
| 2:3:4:5 | /30:/20:/15:/12 | 1.73 |
| 3:4:5:6 | /20:/15:/12:/10 | 2.12 |
| 4:5:6:8 | /30:/24:/20:/15 | 2.48 |

КОНЕЦ ТЕКСТА